\documentclass{sig-alternate}
\usepackage{sparklines}
\usepackage{booktabs}
\usepackage{multirow}
\usepackage[hyphens]{url} 
\usepackage[english]{babel}
\usepackage{array}
\usepackage{subfigure}
\usepackage{chngcntr}

\newcolumntype{C}[1]{>{\centering\arraybackslash}p{#1}}

\begin{document}
\makeatletter
\def\@copyrightspace{\relax}
\makeatother

\title{What They Do in Shadows: Twitter Underground \\Follower Market}
%
%
%
%
%

\numberofauthors{2} 
%
\author{
%
%
\alignauthor
Anupama Aggarwal\\
       \affaddr{IIIT-Delhi}\\
       \email{anupamaa@iiitd.ac.in}
\alignauthor
Ponnurangam Kumaraguru\\
       \affaddr{IIIT-Delhi}\\
       \email{pk@iiitd.ac.in}
}

\maketitle
\begin{abstract}
Internet users and businesses are increasingly using online social networks (OSN) to drive audience traffic and increase their popularity. In order to boost social presence, OSN users need to increase the visibility and reach of their online profile, like - Facebook likes, Twitter followers, Instagram comments and Yelp reviews. For example, an increase in Twitter followers not only improves the audience reach of the user but also boosts the perceived social reputation and popularity. This has led to a scope for an underground market that provides followers, likes, comments, etc. via a network of fraudulent and compromised accounts and various collusion techniques.

In this paper, we landscape the underground markets that provide Twitter followers by studying their basic building blocks - merchants, customers and phony followers. We charecterize the services provided by merchants to understand their operational structure and market hierarchy. Twitter underground markets can operationalize using a premium ~monetary ~scheme or other ~incentivized \emph{~freemium} ~~schemes. We find out that freemium market has an oligopoly structure with few merchants being the market leaders. We also show that merchant popularity does not have any correlation with the quality of service provided by the merchant to its customers. Our findings also shed light on the characteristics and quality of market customers and the phony followers provided. We draw comparison between legitimate users and phony followers, and find out key identifiers to separate such users. With the help of these differentiating features, we build a supervised learning model to predict suspicious following behaviour with an accuracy of 89.2\%. \\
\end{abstract}




\section{Introduction}\label{sec:intro}
Social media presence has become vital for businesses for lead generation, and users to ~increase ~their ~popularity ~~amongst their friends network. In order to enhance and maintain social media presence, users need to generate a following for their social profile, such as - likes on the Facebook page, followers on Twitter and comments on Instagram post. Recent studies have indicated the growth of underground markets for the purchase of Twitter followers, Facebook likes, Instagram followers and Yelp reviews~\cite{insta2014dp, yelp:2014bs, stringhini2012poultry, stringhini2013follow, thomas2013trafficking, viswanath2014towards}. Users subscribe to services of underground markets to artificially boost their social media presence and influence.  

An increase in Twitter followers not only improves the audience reach of the user but also boosts her perceived social reputation and popularity. Rising demand of Twitter followers has led to the growth of an underground industry that caters to users' need for quick followers. We refer to this underground industry as \emph{follower market} and to their operators as \emph{follower merchants}. As per a 2014 study, selling fake Twitter followers generates a revenue up to an estimated \$360 million per year~\cite{twtr:2014fk}. Service provided by the merchants is not restricted to monetary payment by the customers. There exist two major types of operational schemes in Twitter underground markets - \emph{premium} and \emph{freemium}. In \emph{premium} markets, the customer has to pay to the merchant in order to gain followers. \emph{Freemium} markets operate by making the customer authorize merchant's Twitter application, hence including the customer in a phony follower collusion network.

Underground market has constantly evolving techniques to provide phony followers like (i) selling fraudulent accounts, i.e., pseudonym accounts which act as fake followers; (ii) using compromised accounts where the malware on user's machine or compromised credentials cause her to follow customer's account without user's knowledge; (iii) leveraging collusion networks where customers are incentivized to become part of the follower network~\cite{stringhini2013follow, thomas2013trafficking}. To understand the structure and characteristics of Twitter \emph{follower market}, we focus on its basic building blocks viz. (i) \emph{follower merchants}, (ii) \emph{customers}, i.e. the user who take services from these merchants, and (iii) \emph{phony followers}, i.e., the followers provided as a service to a customer by the merchant. An artificially inflated follower count can give the user a veneer of importance and popularity in Twittersphere. In an attempt to disrupt the operations of the Twitter follower market, researchers have proposed techniques to detect customers in paid markets by characterizing their behavioral patterns in contrast to legitimate users~\cite{stringhini2012poultry, stringhini2013follow}. In this study, we do not limit ourselves to only paid markets but also characterize customers of \emph{freemium} markets. Previous studies have also investigated the impact of fraudulent accounts created by the merchants which are used as phony followers, and have developed supervised learning model to detect such Twitter accounts~\cite{thomas2013trafficking}. However, we broaden our study to encompass not only fraudulent accounts, but also the followers who are legitimate users part of the collusion follower network.  

This study landscapes the Twitter follower market. Characterization and analysis of 60 \emph{freemium} and 57 \emph{premium} markets shed light on (i) structure of the follower merchants, (ii) quality of customers and (iii) key identifiers to distinguish between phony follower accounts and legitimate users. In particular, we present following contributions: 

First, we conduct a longitudinal study to characterize Twitter \emph{follower merchants}. In order to identify the most popular merchants and market leaders, we introduce the idea of \emph{Quality of Service} (QoS) for the Twitter follower markets. QoS is an important parameter to judge the overall performance of a service, in our case, the follower merchants~\cite{bolton1991multistage}. As discussed in Section~\ref{subsec:qos}, we define a metric for QoS, which takes into account the services promised by the merchant, expectation by the customer and difference between the two. Using the QoS and popularity metric, we are able to highlight a hierarchy of follower merchants in the underground market and show that this market exhibits an \emph{oligopoly} structure.  \\

Second, we assess the \emph{customers} taking services from the follower market. We characterize customers of various merchants on the basis of their social reputation and profile attributes. We observe that customers lying on the higher strata of quality take services of freemium merchants.   \\

Lastly, We present an anatomy of the purchased Twitter followers. We characterize profile attributes and behavioural features of purchased followers. We identify key indicators to distinguish between suspicious following behaviour from that of legitimate Twitter users. We use these identifiers and build a supervised learning mechanism which detects suspicious following behaviour with an accuracy of 89.2\%.

\section{Related Work}\label{sec:related}
In this section, we present some closely related work focussed on underground markets on Web 2.0. In particular, we summarize previous literature on analysis and detection of social media underground markets.
\paragraph{Social Media Underground Market}
Researchers have shown that miscreants use several strategies to monetize spam and other malicious activities~\cite{levchenko2011click}. There exists a large underground market which sells specialized services and products like fraudulent accounts~\cite{thomas2011suspended, thomas2013trafficking}, solving CAPTCHA ~\cite{motoyama2010re}, pay-per-install~\cite{caballero2011measuring}, and writing fake reviews or website content~\cite{motoyama2011dirty, wang2012serf}. Social media users take such services to increase their online presence. For example, on Twitter, users attempt to gain followers in order to boost their popularity~\cite{cha2010measuring}. Underground markets are a threat to the quality of service and are generating a revenue of about \$360 million  per year from sale of fake Twitter followers~\cite{twtr:2014fk}. In this paper, we present a comprehensive study of Twitter follower market to understand how it operates and assess the Quality of Service~\cite{bolton1991multistage} and percieved gain by the use of phony follower merchants. 

\paragraph{Fraudulent Account Detection}
Recent studies have shown that merchants often create fake accounts to deliver services like phone verified email accounts~\cite{thomasdialing}, Twitter followers~\cite{thomas2013trafficking} and Facebook Likes~\cite{beutel2013copycatch}. Researchers show that such fraudulent accounts can be detected at the time of account creation by merchants by finding patterns in account naming convention and registration process~\cite{thomas2013trafficking}. In this paper, we focus on phony followers of the customers. However, we do not limit our study to only fraudulently created accounts; we charecterize and detect fake as well as legitimate accounts that exhibit suspicious following behaviour.

\paragraph{Merchant and Customer Detection}
Social media users are increasingly taking underground market services to increase their following. Researchers have modelled suspicious Twitter following behaviour by identifying difference in follow pattern from the majority~\cite{jiang2014detecting}. Previous studies highlight the unfollow dynamics of the victim customer accounts whose credentials are compromised by merchants~\cite{stringhini2013follow}. In this paper, we study not only compromised users but also the legitimate users part of the collusion follower network. 

\section{Background}\label{sec:background}  
The goal of our study is to characterize the underground follower market. Before we elaborate our dataset and methodology, in this section, we briefly describe the popular market schemes. We also discuss the QoS metric to quantify the quality and uncover the underlying hierarchy of merchants.

\subsection{Freemium and Premium Markets}\label{subsec:frpr}
The underground follower market operates by either selling followers or incentivizing the customers to become part of the phony follower network in order to increase follower count. Here, we present in brief how \emph{freemium} and \emph{premium} market schemes operate.

\paragraph{\textbf{Premium Market}} Under the \emph{premium} scheme, customers have to pay money to the merchants in order to gain followers. Customer provides a Twitter username to the merchants for which she wants to increase the follower count, and buys a specific package (e.g. package - 1000 followers for \$3). Within a span of few minutes to few hours, depending upon the merchant, follower count of the specific Twitter username (provided to the merchant) increases as per the package. Premium market allows a customer to buy bulk followers for not just himself but any Twitter user. This has opened up new possibilities of exploitation by spammers. In 2014, a group of hoaxters known for their notoriety, flooded a prominent online daily's Twitter account with fake followers to damage the brand~\cite{hoaxters:2014qy}. In this study, we analyse the structure and characteristics of 57 such merchant websites which sell bulk followers. The merchants under the premium scheme, either simply sell followers in exchange of money, or also require the customer to provide her Twitter account's password so that the customer can be made part of the collusion phony follower network. Note that in both the cases, merchants require monetary payment from the customer.

\paragraph{\textbf{Freemium Market}} \emph{Freemium} market scheme lets the customer gain followers without any monetary payment. However, in return, the customer needs to authorize merchant's Twitter application that enables the merchant to include customer in the collusion phony follower network. Most of the freemium merchants display their recent customers on their website and keep refreshing the list after every few minutes. Once the customer authorize merchant's app, she starts gaining followers within minutes. The merchant app includes various permissions like - \emph{see who you follow, follow new people}, \emph{update your profile}, and \emph{post tweets for you}. These permissions enable the merchant to make the customer follow other Twitter users (which may be other customers of the merchant) and also post promotional tweets on her behalf. Since the customer does not have to provide her password unlike the second scheme in premium markets as discussed above, the customer is at a much lower risk of being compromised. 

\begin{figure}[h]
  \centering
    \includegraphics[width=0.4\textwidth]{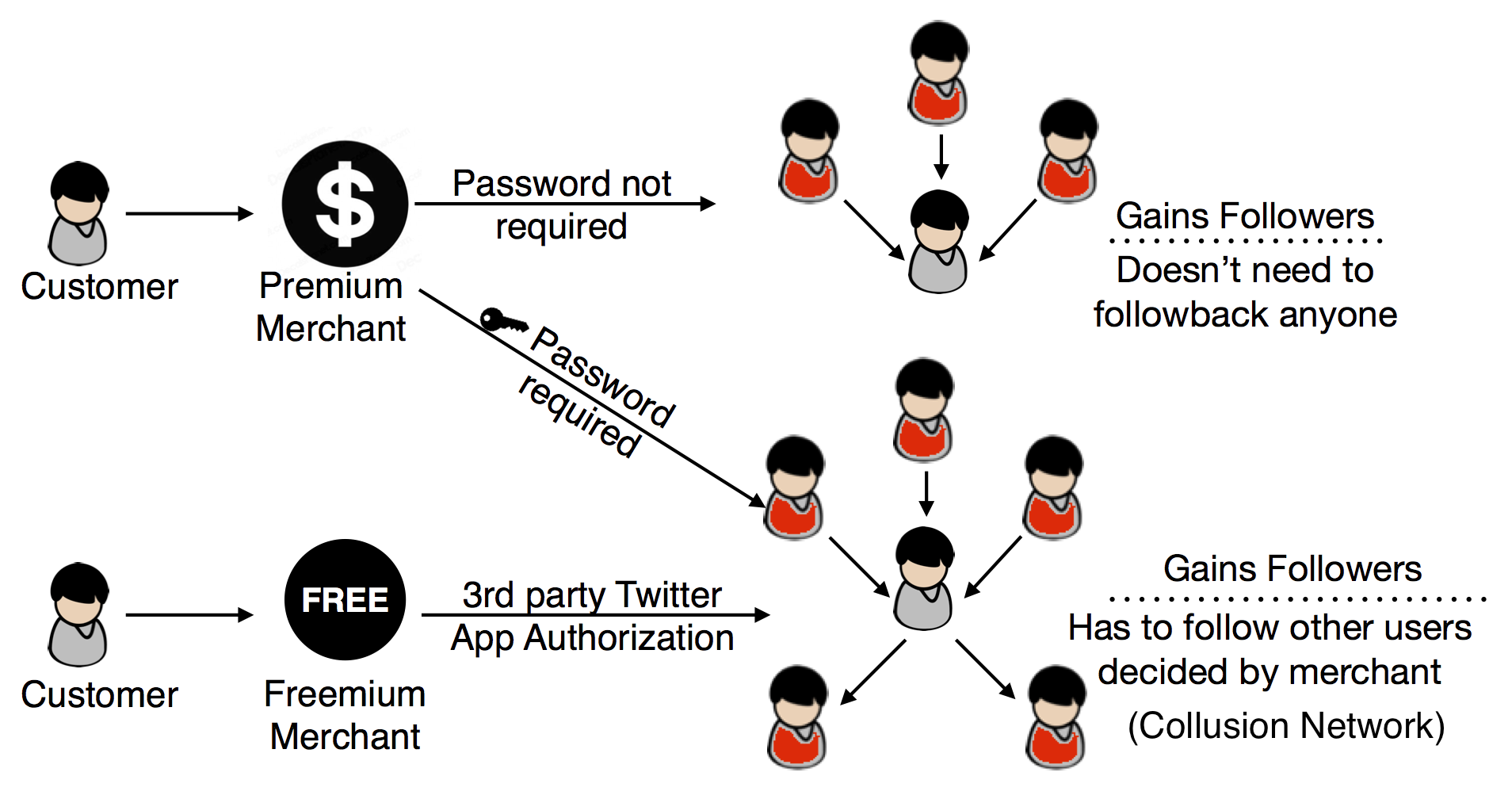}
    \caption{\scriptsize{Different market schemes of follower market. Merchants can enforce monetary payment or use Twitter application to incentivize the customer. In both schemes, merchant can leverage collusion networks to include customers into the phony follower network.}}\label{fig:FrPrMarkets}
\end{figure}

Figure~\ref{fig:FrPrMarkets} summarizes how the two market schemes operate. It can be seen that freemium market operate primarily by leveraging collusion networks. This causes even legitimate users to exhibit phony follow behavior in return of bulk followers. Customers who provide their passwords to merchants under the premium scheme are at a high risk of being compromised. Once compromised, these accounts can be fully controlled by the merchants and used for following other customers, spamming or sending promotional tweets as and when required.   

\subsection{Quality of Service}\label{subsec:qos}
Twitter follower merchants lay down various terms of services to the customers who subscribe to them. As of now, we do not understand to what extent these merchants violate their terms of services, who are the market leaders and which merchants have maximum penetration in the market. To answer all these questions, we use existing literure on consumer research by Bolten et al. which provides a conceptual framework to model customer's assessment of service quality and value~\cite{bolton1991multistage}. We apply the formulization of Quality of Service (QoS) to the underground follower merchants to understand the quality and hierarchy of merchants. 

Researchers have shown that the measure of QoS can be based on performance of the service, expectation by the customer and the gap between these two parameters. Based on this, QoS can be defined as a function \emph{q} --

\small
\begin{equation}
\textbf{QoS} = q(PERFORM, EXPECT, DISCONFIRM)
\end{equation}
\normalsize
where PERFORM is a vector of the performance of a merchant based on several terms of services \{$SA_1$...$SA_k$\}, i.e.,

\small
\begin{equation}
PERFORM= p_k(SA_k)
\end{equation}
\normalsize
EXPECT is a vector which describes the prior expectations of the customer for each term of service $\in$ \{$SA_1$...$SA_k$\}. DISCONFIRM is a vector describing the amount of discrepancy between performance of the merchant and expectation of the customer for each of the terms of services. 

This definition of QoS helps us to evaluate the performance of the underground merchants and understand the structure of Twitter follower market. In Section~\ref{subsubsection:quality}, we describe in detail how we use this QoS formulization to assess the performance of merchants.

\section{Dataset and Methodology}\label{sec:dataset}
In this section, we describe our dataset and explain the methodology we use for the analysis of underground market. 

\subsection{Data Collection Methodology}\label{subsec:collection}
We collect data from both \emph{premium} as well as \emph{freemium} merchants. Since we want to landscape the merchants, customers and followers, we undergo a three step process to collect data from freemium and premium merchants.

\paragraph{\textbf{Merchant Identification}}
One of the building blocks of underground follower market is the \emph{merchants}, i.e., the operators who provide phony followers to customers. In order to identify the merchant websites, we use search engine queries and filter results of Twitter search. Table~\ref{tb:keywords} shows the list of keywords which we use to identify merchant websites from premium and freemium market. We limit our search space to web engine (Google and Bing) results, tweets and Twitter user profile descriptions which match the keywords or a combination of keywords in the category.

\begin{table}[h]
\caption{\scriptsize{shows keywords which are used to identify merchants. Search engine results are directly used after manual filtering. Twitter search results reveal promotional tweets as well as user profiles which reveal more merchant websites.}}\label{tb:keywords}
\scriptsize
\begin{tabular}{@{}cC{3cm}C{3cm}@{}}
\toprule
 & \multicolumn{2}{c}{\textbf{Keywords}} \\ \midrule
 & \textbf{Premium} & 	\textbf{Freemium} \\ \midrule
\textbf{Web Search} & buy, Twitter, followers, bulk, increase, order, grow, gain, more, real, cheap & latest, riders, free, followers, Twitter, `get more' \\ \midrule
\begin{tabular}[c]{@{}c@{}}\textbf{Twitter}\\ \textbf{Search}\end{tabular} & \multicolumn{2}{C{6cm}}{recommend to gain, gain followers, gamer follow train, wana gain followers, need more followers, cheap followers} \\ \bottomrule
\end{tabular}
\end{table}

\begin{figure}[h]
\centering
\subfigure[\scriptsize{Merchants located by Twitter search and web engine search in each category.}]{\includegraphics[width=0.23\textwidth]{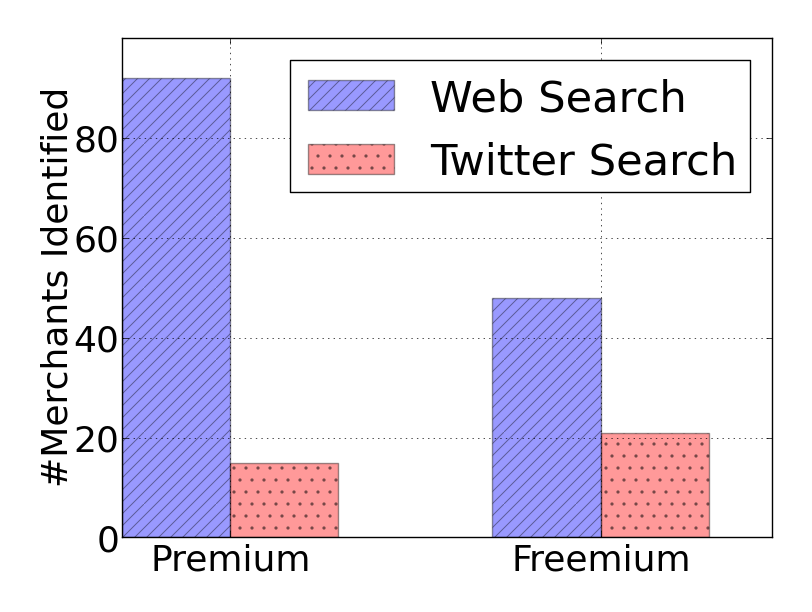}}\label{fig:searchmerchant}
\hspace{2pt}
\subfigure[\scriptsize{Pr=premium, Fr=freemium and Tw=merchants which have a Twitter profile.}]{\includegraphics[width=0.23\textwidth]{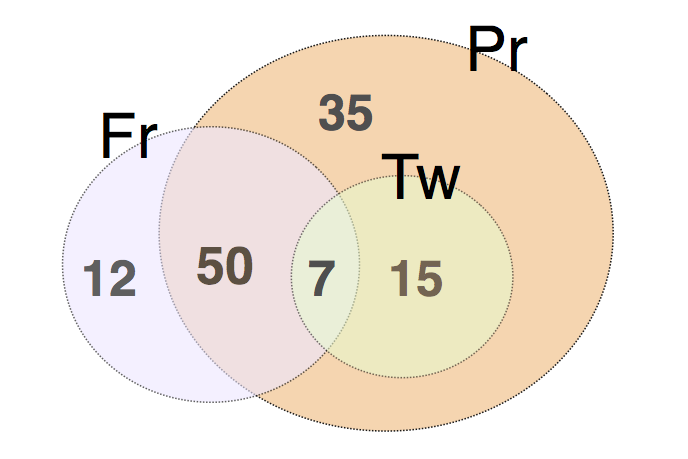}}\label{fig:venndistrib}
\caption{\scriptsize{Distribution of premium and freemium merchants in our dataset. We use Twitter search and web search queries to locate a merchant. We find that some merchants have a Twitter profile and offer both premium and freemium services.}}
\label{fig:datadistrib}
\end{figure}

We manually clean the web search results to identify merchant websites and group them into two categories - `freemium' and `premium'. In case of Twitter search, there was little difference between the promotional tweets and profile descriptions of freemium and premium merchants. Therefore, we use the same set of keywords to identify them and manually group them once we obtain the merchant website URLs. Some merchants offer both \emph{freemium} and \emph{premium} schemes, therefore, they are grouped into both the categories. Figure~\ref{fig:datadistrib}(a) shows the distribution of websites we obtain by Twitter search and web engine queries. Freemium merchants have a larger tweet presence than premium merchants. We posit that this is because freemium merchants primarily operate using collusion network and sending promotional tweets. Twitter search also reveals that few merchants maintain a Twitter profile to gain audience. Figure~\ref{fig:datadistrib}(b) shows the distribution of merchants in both categories and indicates the merchants who maintain a Twitter profile in our dataset. We identify 69 \emph{freemium} merchants and 107 \emph{premium} merchants. Some of these merchants provide both freemium and premium schemes as indicated in the figure. 50 merchants provided both schemes but did not have a Twitter profile.
Out of the 69 freemium merchants we identify, only 60 merchant websites publically display the latest customers on their websites. Also, we buy followers from 57 of the premium merchants in our dataset. Next, we describe in detail how we obtain the dataset of phony followers. Though we were able to identify over 100 merchants, we use the aforementioned dataset because (i) some of the merchant websites did not let us scrape them to collect their customer Twitter handles, (ii) some of the payment gateways did not seem to be safe and hence we did not make a purchase due to security reasons.

\paragraph{\textbf{Phony Follower Data Collection}}
To identify large number of phony followers, we subscribed to services of premium and freemium merchants. For this task, we created dummy Twitter accounts to get followers from freemium and premium merchants. Table~\ref{tb:buyingdistrib} shows services provided by the merchants from where we receive the phony followers. 

In case of freemium merchants, we used dummy accounts to subscribe to each of the merchants; we used one dummy account per merchant. We authorized the Twitter application of each merchant website, in return of which we recieved followers. The followers were added to our dummy accounts at an average rate of 84 unique followers per hour as shown in Table~\ref{tb:dataset} and overall we received 82,808 followers. For premium merchants, we used one dummy account per merchant to gain followers. We purchased the basic package from each of the merchants and obtained 87,458 followers. Table~\ref{tb:buyingdistrib} shows that in our dataset we have 5 merchants which offer only premium services. Overall, we recieved 170,356 followers which wr consider have a suspicious following behaviour. Out of these 5 merchants, 2 merchants did not require customer password as part of any of their follower packages. Therefore, we assume that the followers obtained from these 2 merchants are not part of any collusion network and are hence either fraudulently created or compromised accounts.

\begin{table}[h]
\caption{\scriptsize{Shows the types of services provided by the merchants from where we recieved our dataset of phony followers.}}\label{tb:buyingdistrib}
\scriptsize
\begin{tabular}{@{}lccc@{}}
\toprule
 & \textbf{Model} & \textbf{\#Merchants} & \textbf{\#Followers} \\ \midrule
\multirow{2}{*}{\textbf{Freemium}} & Only Freemium & 12 & \multirow{2}{*}{82,808} \\
 & Freemium + Premium & 57 &  \\ \midrule
\multirow{2}{*}{\textbf{Premium}} & Only Premium & 5 & \multirow{2}{*}{87,458} \\ 
 & Freemium + Premium & 52 &  \\ \bottomrule
\end{tabular}
\end{table}

We ensured that we subscribed to a merchant as soon as we created the dummy Twitter account for that merchant. Therefore, we assume that any follower gained by the dummy account is part of the phony follower network provided by the subscribed merchant. As shown in Table~\ref{tb:dataset}, we further observe that there is a lot of variation in number of followers obtained in case of freemium market, though we subscribed to similar services. Likewise, in case of premium market, we ordered minimum 1,000 followers but received as low as 738 followers. This indicates clear difference in operations of various merchants, which we further discuss in Section~5.1.1.

\begin{table}[h]
\caption{\scriptsize{Shows the number of phony followers collected from each kind of market. The last column shows the distribution of each parameter for all the merchants of specified category.}}\label{tb:dataset}
\scriptsize
\begin{tabular}{@{}lllllll@{}}
\toprule
 &  & Mean & Median & Min & Max & Distribution \\ \midrule

\multirow{2}{*}{Fr} & Followers/Hr & 84 & 85 & 60 & 105 & 
\begin{sparkline}{9}
\spark 	0.018	0.60
		0.036	0.60
		0.054	0.62
		0.072	0.64
		0.090	0.65
		0.109	0.65
		0.127	0.65
		0.145	0.67
		0.163	0.68
		0.181	0.70
		0.2		0.75
		0.218	0.75	
		0.236	0.75
		0.254	0.75
		0.272	0.76
		0.290	0.78
		0.309	0.79
		0.327	0.79
		0.345	0.80
		0.363	0.80
		0.381	0.80
		0.4		0.80
		0.418	0.80	
		0.427	0.81
		0.436	0.85
		0.454	0.85
		0.490	0.85	
		0.509	0.85
		0.527	0.89
		0.545	0.90
		0.563	0.90
		0.581	0.90
		0.6		0.90
		0.618	0.90	
		0.636	0.91
		0.654	0.92
		0.672	0.92
		0.690	0.93
		0.709	0.93
		0.727	0.94
		0.745	0.95
		0.763	0.95
		0.781	0.95
		0.8		0.96
		0.818	0.96	
		0.836	0.96
		0.854	0.98
		0.872	0.99
		0.890	1.0
		0.909	1.0
		0.927	1.0
		0.945	1.0
		0.963	1.0
		0.981	1.02
		1		0.05 /
\end{sparkline} \vspace{8pt} \\

 & Followers & 1,505 & 1,524 & 678 & 2,030 & 
\begin{sparkline}{9}
\spark 0.018	0.678
0.036	0.876
0.054	0.899
0.072	0.912
0.090	0.998
0.109	1
0.127	1.003
0.145	1.025
0.163	1.034
0.181	1.113
0.2	1.216
0.218	1.287
0.236	1.296
0.254	1.298
0.272	1.321
0.290	1.325
0.309	1.34
0.327	1.354
0.345	1.429
0.363	1.432
0.381	1.462
0.4	1.5
0.418	1.502
0.427	1.509
0.436	1.511
0.454	1.519
0.490	1.523
0.509	1.524
0.527	1.528
0.545	1.542
0.563	1.548
0.581	1.56
0.6		1.561
0.618	1.562
0.636	1.562
0.654	1.652
0.672	1.671
0.690	1.673
0.709	1.698
0.727	1.723
0.745	1.762
0.763	1.8
0.781	1.803
0.8		1.812
0.818	1.845
0.836	1.865
0.854	1.902
0.872	1.958
0.890	1.968
0.909	1.969
0.927	1.979
0.945	1.98
0.963	1.982
0.981	1.987 /
\end{sparkline} \\ \midrule

\multirow{2}{*}{Pr} & \begin{tabular}[c]{@{}l@{}}Cost/1000\\ Followers\end{tabular} 

& \$8.4 & \$8 & \$3 & \$14 & 
\begin{sparkline}{9}
\spark
0.018	0.3
0.036	0.3
0.054	0.5
0.072	0.6
0.090	0.6
0.109	0.6
0.127	0.6
0.145	0.6
0.163	0.6
0.181	0.7
0.2	0.7
0.218	0.7
0.236	0.7
0.254	0.8
0.272	0.8
0.290	0.8
0.309	0.8
0.327	0.8
0.345	0.8
0.363	0.8
0.381	0.8
0.4	0.8
0.418	0.8
0.427	0.8
0.436	0.8
0.454	0.8
0.490	0.8
0.509	0.8
0.527	0.9
0.545	0.9
0.563	0.9
0.581	0.9
0.6	0.9
0.618	0.9
0.636	0.9
0.654	0.9
0.672	0.9
0.690	0.9
0.709	0.9
0.727	0.9
0.745	0.9
0.763	0.9
0.781	0.9
0.8	1
0.818	1
0.836	1
0.854	1
0.872	1
0.890	1
0.909	1
0.927	1
0.945	1
0.963	1.4
0.981	1.4
1	1.4 /
\end{sparkline} \vspace{2pt} \\

 & Followers & 1,590 & 1,607 & 738 & 2,095 & 
\begin{sparkline}{9}
\spark
0.018	0.738
0.036	0.974
0.054	0.998
0.072	1.012
0.090	1.06
0.109	1.098
0.127	1.103
0.145	1.125
0.163	1.134
0.181	1.209
0.2	1.312
0.218	1.351
0.236	1.388
0.254	1.39
0.272	1.411
0.290	1.419
0.309	1.42
0.327	1.446
0.345	1.521
0.363	1.534
0.381	1.555
0.4	1.565
0.418	1.589
0.427	1.59
0.436	1.597
0.454	1.603
0.490	1.604
0.509	1.607
0.527	1.626
0.545	1.632
0.563	1.637
0.581	1.638
0.6	1.639
0.618	1.646
0.636	1.652
0.654	1.731
0.672	1.748
0.690	1.752
0.709	1.778
0.727	1.816
0.745	1.837
0.763	1.862
0.781	1.883
0.8	1.907
0.818	1.935
0.836	1.937
0.854	1.967
0.872	2.044
0.890	2.048
0.909	2.048
0.927	2.049
0.945	2.059
0.963	2.067
0.981	2.072
1	2.095 /
\end{sparkline} \\ \bottomrule
\end{tabular}
\end{table}

\paragraph{\textbf{Market Customer Ground Truth Data}}
Customer users of premium merchants are not disclosed by the merchant websites. However, in case of freemium merchants, the latest customers are often displayed on the website with a link to their Twitter profile, and the list is refreshed after every few minutes. In order to collect the ground truth data of customers, we scrape 60 such freemium merchant websites which provide link to their latest customers and obtain 171k unique customer profiles by taking hourly snapshots. Table~\ref{tb:cust} describes the customer dataset in more detail.

\begin{table}[h]
\center
\caption{\scriptsize{Dataset description of customers located in freemium market over a period of 4 months by scraping 60 merchant websites.}}\label{tb:cust}
\scriptsize
\begin{tabular}{ll}
\toprule
Data Collection Timeframe   & 2014-07-18 - 2014-11-18 \\
Number of Snapshots         & 2,870                    \\
Number of Merchants         & 60                      \\
Number of Customers Located & 171,234                 \\
Verified Customers          &  10 					\\ \bottomrule                      
\end{tabular}
\end{table}

Customers of premium merchants are not disclosed. Hence, we limit our study to the customers of freemium market and study their behaviour.

\subsection{Ethical Consideration}\label{subsec:ethical}   
We ensured that all money we paid to underground merchants to acquire fake followers was exclusively for the dummy Twitter accounts we created. The dummy accounts were fully controlled by us and were for the sole purpose of conducting experiments in this paper. We revoked the merchant's application after the experiments were conducted to ensure that we do not cater to the collusion follower network.

\section{Landscaping Twitter Follower Underground Market}\label{sec:landscape}
In this section, we study in detail the characteristics of the merchants, customers and followers provided by the merchants in deatil, to understand the structure of underground follower market.

\subsection{Twitter Follower Merchants}\label{subsec:merchants}
Merchants are the market operators which provide phony followers to their customers. We recall that merchants can offer \emph{premium}, \emph{freemium} or both the schemes to their customers. 

\subsubsection{Merchants violate their promises}\label{subsubsection:violation}
The merchants of underground Twitter follower market offer various guaranteed services to customer at the time of subscription. Many merchants promise services like authentic followers, moneyback guarantees, quick followers and follower retention which encourage the customers to either buy bulk followers from premium merchants or subscribe to freemium merchants. Table~\ref{tb:promises} shows the list of most common promises made by the merchants. Though these promises and guarantees seem lucrative, and hence attract a lot of customers, merchants often violate these services.  

\begin{figure}[h]
  \centering
    \includegraphics[width=0.3\textwidth]{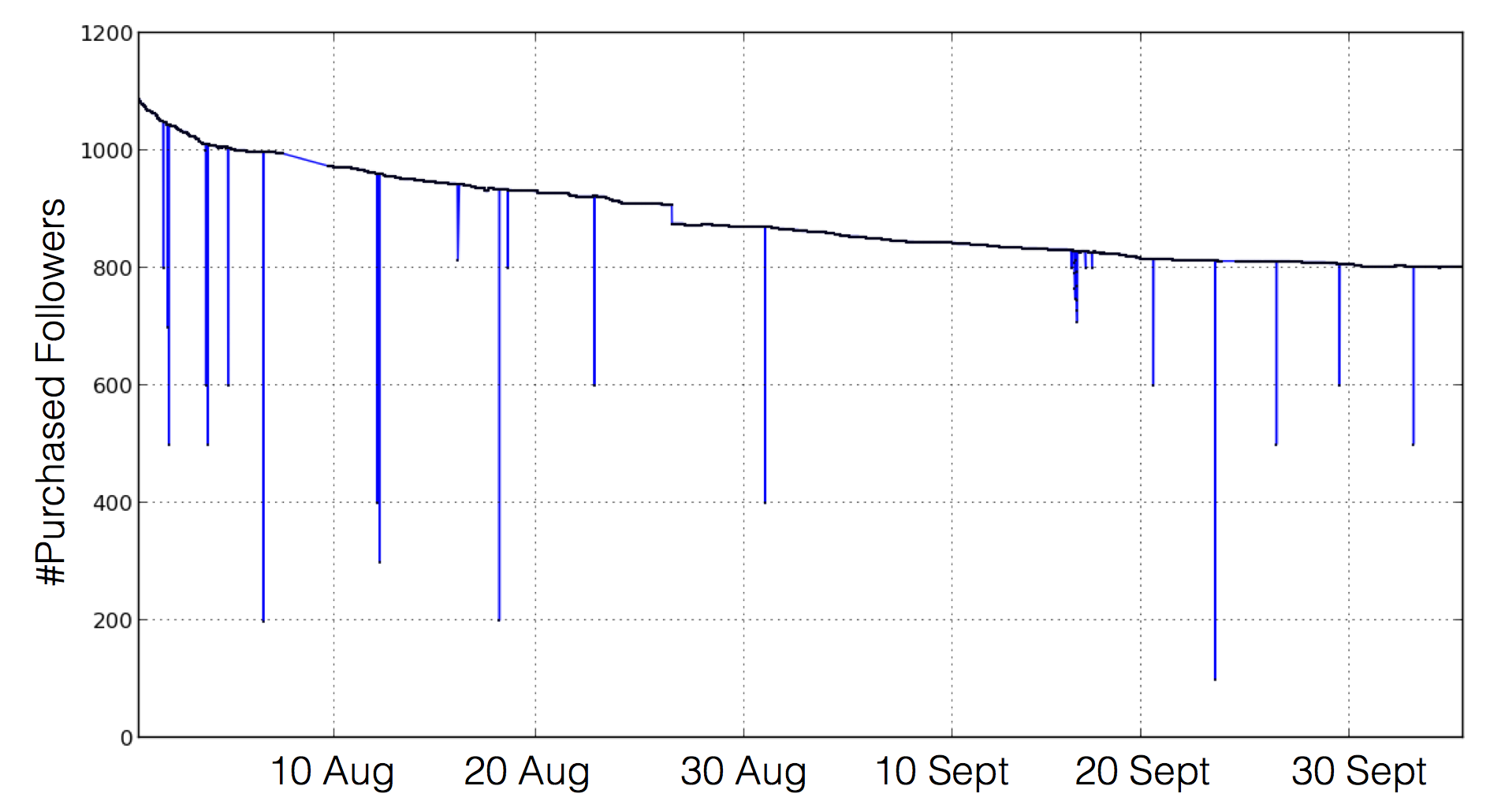}
    \caption{\scriptsize{Shows the dips in follower count every hour for one of the most prominent premium merchants in our dataset. Few minutes/hours after the dips, the follower count rises back.}}\label{fig:followerdips}
\end{figure}

\paragraph{Lack of follower retention}
Most of the preemium merchants provide follower retention policy, i.e., they state that - \emph{``if you loose any number of followers...we'll refill the page with the lagging followers, at absolutely free of cost"}. Therefore, we expect that customers will always have the same number of followers at any point of time which they initially purchased. To assess whether this is true or not, we continuously monitor the followers gained by our dummy customer accounts for the premium market. We observed that only 3 merchants provided us lesser followers than promised and rest 54 merchants provided us either the exact number of requested followers or more (as previously seen in Table~\ref{tb:dataset}). However, after the date of purchase and the gain of requested amount of followers, we observed constant drop in the follower count in case of all the customers. We further investigated the drop in follower count by taking hourly snapshots of each of the customer profile and observed constant fluctuations in the follower count during each day. Figure~\ref{fig:followerdips} shows this phenomena for one of the popular premium merchants in our dataset. 

To analyse whether the dips in follower count are at a specific time, we measured correlation of follower count with hour of the day. We calculated the Pearson Correlation Coefficient (PCC) between the follower count distribution and the corresponding hour of the day of snapshot time. We found that there does not exist any correlation between the follower count and time for any of the merchants (max PCC = 0.02, indicating negligible correlation). We also observed that there were several dips in the follower counts in both markets, however, we did not gain new users as followers. Users from the same set of initially obtained users kept unfollowing and following us back. We posit that the reason behind such following behaviour is that since the follower accounts' activities are controlled by the merchants, they optimize who to follow to cater to a larger customer base without raising suspicion of following too many users at a given point of time.

\begin{table}[h]
\center
\scriptsize
\caption{\scriptsize{List of most popular promises and guarantees made by the merchant to customers.}}\label{tb:promises}
\begin{tabular}{@{}cl@{}}
\toprule
\multicolumn{1}{l}{}      & \textbf{Promises by Merchants to Customers} \\ \midrule
\multirow{3}{*}{\textbf{Freemium}} & \emph{60+ new followers per ride}        \\
                          & \emph{promotional status updates}         \\
                          & \emph{never alter profile information}     \\  \midrule
\multirow{5}{*}{\textbf{Premium}}  & \emph{new followers every minute}         \\
                          & \emph{Ad free} -- No promotional tweets      \\
                          & \emph{3 month no drop guarantee} -- follower retention policy     \\
                          & \emph{genuine profile}  -- legitimate users will follow          \\
                          & \emph{delivered in 1-2 days}              \\ \bottomrule
\end{tabular}
\end{table}

\subsubsection{Quality evaluation of merchants}\label{subsubsection:quality}
\begin{figure}[ht]
  \centering
    \includegraphics[width=0.45\textwidth]{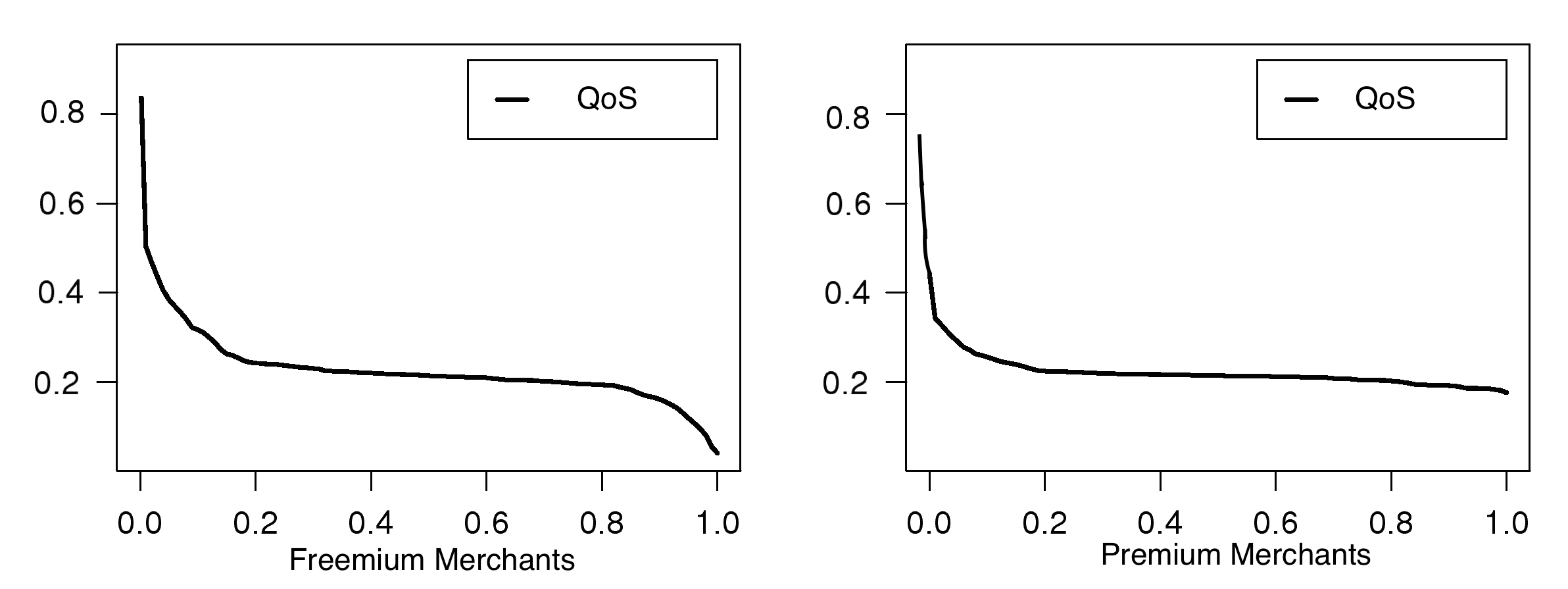}
    \caption{\scriptsize{Quality of Service of Freemium and Premium Markets.}}\label{fig:QoSFrPr}
\end{figure}

Catering to the needs of customer is the key to successful busineses. Twitter follower merchants make several promises to attract and retain customers, as previously indicated in previous section. Therefore, we calculate Quality of Service of the merchants to understand how well they are catering to their customers. We use the following definition of QoS and its parameters as described earlier in Equation~(1)
\small
$$
\textbf{QoS} = q(PERFORM, EXPECT, DISCONFIRM)
$$
\normalsize
For simplicity, we give equal weightage to all the promises, and hence $p_k$ = 1 in Equation (2). DISCONFIRM is the discrepency between performance of merchant and expecation of the customer for a particular promise made by the merchant. For a specific merchant, we calculate the collective QoS based on all the promises \{$SA_1$...$SA_N$\} which it provides as - 
\small
$$
QoS = \frac{\sum_{i=1}^N \bigg[ 1 - \Big( \frac{EXPECT_i - PERFORM_i}{PERFORM_i} \Big) \bigg]}{N} 
$$
\normalsize
Note that in case a merchant overdelivers a certain promise, then the above formulization of QoS for that specific promise gets a value > 1, hence rewarding the merchant. This gives us a normalized value of QoS for all the merchants. Figure~\ref{fig:QoSFrPr} shows the QoS curve for both freemium and premium merchants. The knee point of the QoS curve for freemium market lies at X=0.1, Y=0.3; this indicates that 90\% freemium merchants have a QoS value of 0.3 or less. The knee point for premium merchants is at X=0.05, Y=0.28 indicating that 95\% premium merchants in our dataset have a normalized QoS value of 0.28 or lesser. The highest QoS for freemium market is 0.82, whereas for premium we found it to be 0.78. This shows that overall, QoS for freemium market is higher than that of freemium market. We further investigate and find that the violation of \emph{no drop guarantee} is prime reason behind the low QoS for premium market. We had earlier seen this phenomena of follower drop in Figure~\ref{fig:followerdips}. Frequent drops in follower can raise a red flag against the customer, and hence the merchants which deliver followers exhibiting such phenomena are penalized in our formulization of QoS.

\subsubsection{Few merchants are market leaders}\label{subsubsection:leaders}
Now we evaluate which merchants attract the highest share of user base. In order to do so, we compute the popularity of each merchant by using two metrics - Alexa ranking of merchant website, and number of promotional tweets.~\footnote{Alexa -- \url{http://www.alexa.com/}}

\paragraph{Alexa Ranking} Alexa rank measures  website's popularity based on the traffic to that website. For each merchant, we extract the global rank of its website and then compute the normalized Alexa rank for merchant $M_i$ as followed -- 
\small
$$
Alexa\_Norm_{i} =  1 - \frac{AlexaRank_{i}}{\max\limits_{i \in \{1...N\}}(AlexaRank_{i})}
$$
\normalsize
where N is the total number of merchants. This gives us a normalized measure of website traffic between 0 and 1.

\begin{figure}[ht]
  \centering
    \includegraphics[width=0.35\textwidth]{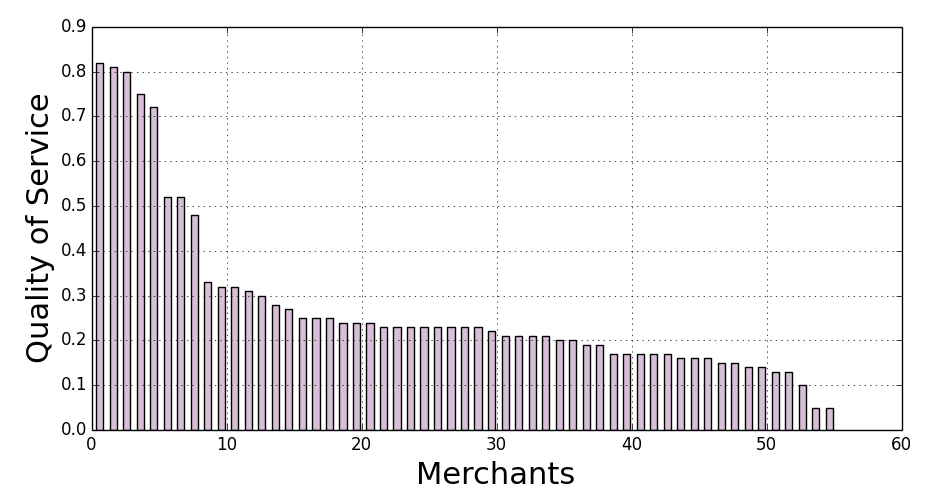}
    \caption{\scriptsize{Quality of Service of Freemium and Premium Markets.}}\label{fig:oligopoly}
\end{figure}

\paragraph{Social Media Popularity} We also use social media popularity of the merchant website by collecting the promotional tweets advertising the merchant if any. We search for each merchant's URL and its Bitly shortened version using Twitter search API. We then define the OSN popularity for each of the N merchants as --
\small
$$
OSN\_Popularity_{i} =  \frac{NumTweet_{i}}{\max\limits_{i \in \{1...N\}}(NumTweet_{i})}
$$
\normalsize
where $NumTweet_i$ is the number of promotional tweet for merchant $M_i$.
Using these two metrics, we finally draw the overall popularity of a merchant website by taking an average of normalized Alexa ranking and OSN Popularity.
\small
$$
Merchant\_Popularity_i = \frac{Alexa\_Norm_{i} +OSN\_Popularity_{i}}{2}
$$
\normalsize
Figure~\ref{fig:oligopoly} shows the distribution of popularity of all the merchant websites. 
We notice that 5 of the merchants have very high popularity as compared to other merchant websites. The top 5 merchants have a normalized popularity score of more than 0.71 (71\%). This indicates that there exists an \emph{oligopoly} hierarchical structure amongst the merchants.~\footnote{Oligopoly: \url{http://www.economicsonline.co.uk/Business_economics/Oligopoly.html}} That is, there are few market leaders which attract most of the audience. We notice that 4 of the market leaders have Twitter verified customers in our dataset.~\footnote{Twitter vefiried users \url{https://support.twitter.com/articles/119135-faqs-about-verified-accounts}} The data collection from freemium merchant websites show that atleast 10 Twitter verified accounts subscribed to services of one of the top 4 merchants represented in Figure~\ref{fig:oligopoly}. This also indicates that market leaders attract high profile customers. 

\subsubsection{Merchant popularity does not reflect quality}\label{subsubsection:opvsqos}
\begin{figure}[ht]
  \centering
    \includegraphics[width=0.3\textwidth]{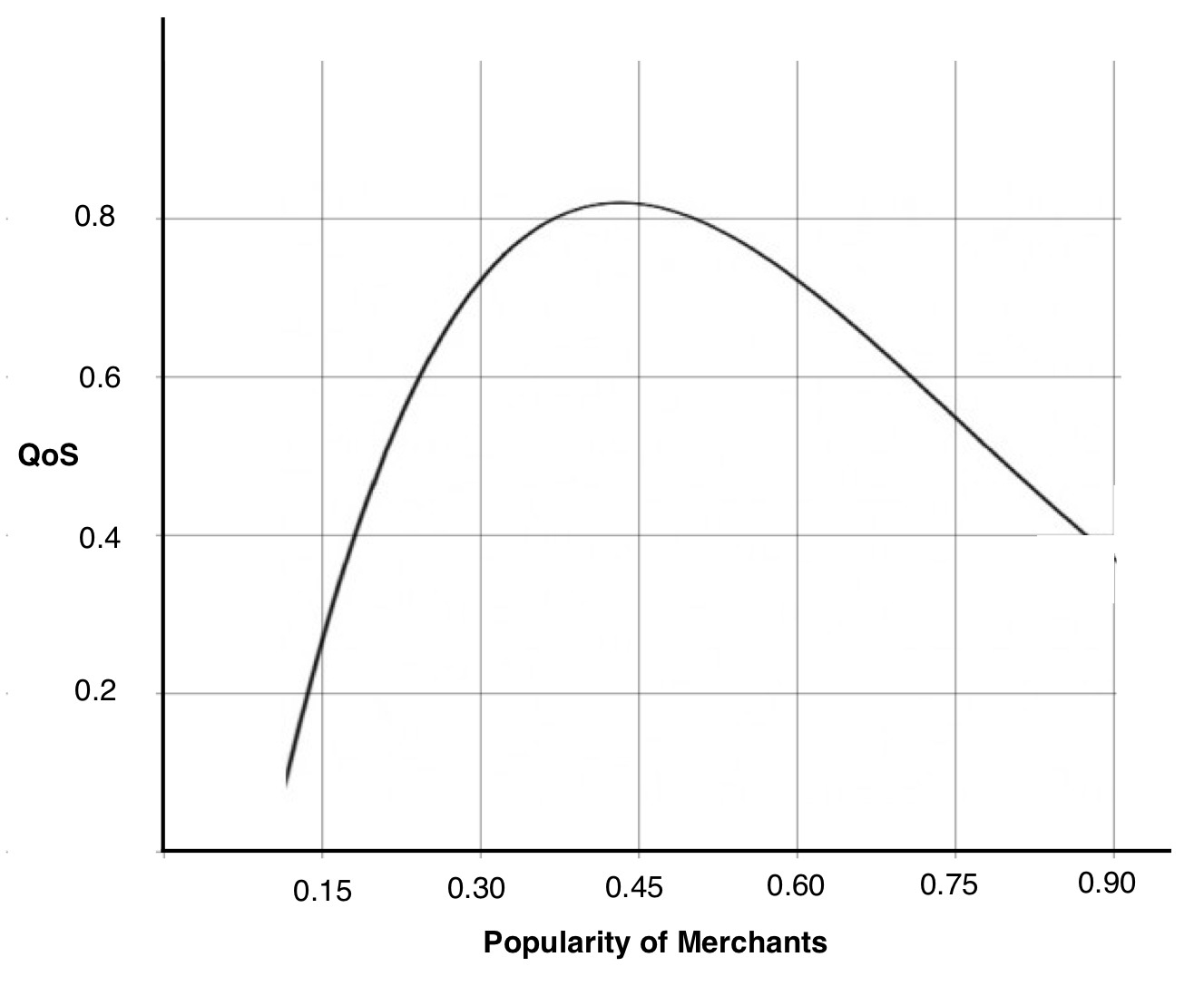}
    \caption{\scriptsize{Quality of Service vs Popularity of Merchants.}}\label{fig:QoS_Popularity}
\end{figure}

Next we try to answer whether the merchant popularity is reflective of its Quality of Service or not. We find that there does not exist any correlation between the QoS and popularity distribution of the merchants. Figure~\ref{fig:QoS_Popularity} shows the distribution of QoS against popularity for the follower merchants. We observe that the most popular merchant has a QoS value of only 0.40. Whereas, the highest quality merchant has a popularity score of 0.42. In order to answer why some merchants with high QoS have low popularity we look at the pricing details and follower-gain/hour of those merchants. We find that for most of such merchants, either the cost of followers is high (in case of premium market) or the follower gain per hour is low (for freemium market). Hence, we postulate that higher cost and lower follower gain must be the reason behind low popularity of these promise delivering, high quality merchants.

\subsection{Underground Market Customers}
Now we study the characteristics of customers who subscribe to follower merchants, to understand who these users are and which services they subscribe to. Customers of premium market are undisclosed, hence, we limit our study to merchants which provide either freemium service, or both freemium and premium service.
\subsubsection{Spammers, wannabes and celebrities}
\begin{figure}[ht]
  \centering
    \includegraphics[width=0.4\textwidth]{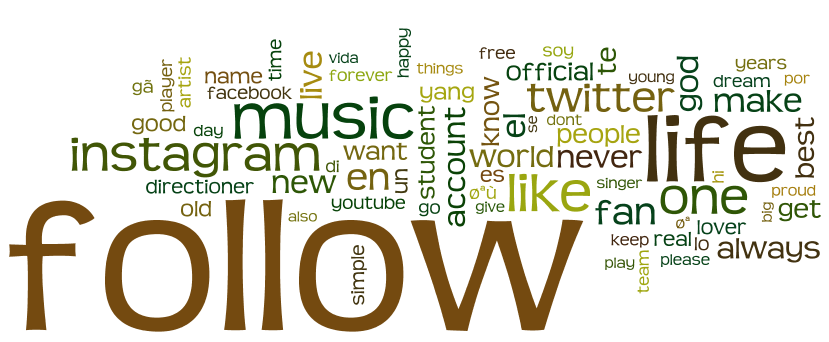}
    \caption{\scriptsize{Wordcloud of bio of the customer profiles.}}\label{fig:wannabes}
\end{figure}

To understand who are the customers of follower market, we use our dataset of 171,234 customers collected by scraping the merchant websites. We find out whether they are listed, verified and analyse their \emph{bio}. Figure~\ref{fig:wannabes} shows that many customers use \emph{`follow'}, \emph{`artist'}, \emph{`director'}, \emph{`music'} in their bio. This indicates that these users are probably \emph{wannabe} artists and are trying to attract a large following. We also analyzed the URLs posted by these users in their latest 200 tweets and found that 10\% of the users had posted atleast one or more URL blacklisted by Google Safebrowsing~\footnote{Google Safebrowsing \url{https://developers.google.com/safe-browsing/}} or Phishtank.~\footnote{Phishtank \url{www.phishtank.com}} We also noticed that 10 of the customers we acquired in our dataset were Twitter verified users and had posted atleast one promotional tweet (which was soon deleted) about the merchant website. This shows that Twitter verified accounts, i.e., celebrity accounts, also use freemium merchants to boost their follower count.

\subsubsection{Market leaders attract prominent customers}
To identify the prominent customers, we use `Klout'~\footnote{\url{http://www.klout.com}} score. `Klout' is a popular tool to measure influence based on various factors like followers, freinds, retweets and favourites. The average Klout score for the social media users is 40.~\footnote{\url{http://support.klout.com/customer/portal/articles/679109-what-is-the-average-klout-score}} We found that 30\% customers had a Klout score of more than 40. Out of these prominent users, 81.7\% users subscribed to atleast one merchant with a Popularity\_Score > 0.72. This indicates that the merchants which are more popular and market leaders attract prominent customers who have a higher reputation. We argue that the vice-versa is not true because we observe that overall 54.3\% customers who had a less than average Klout score subscribed to atleast one of the market leaders. This shows that its not the merchants which are boosting the Klout score of their customers; high reputation users are subscribing to market leaders. 
\subsection{Phony Followers of Underground Market}\label{subsec:followers}
In this section, we present an anatomy of the phony followers which we recieve from various merchants. We find key identifiers which can be helpful to distinhuish from legitimate users and hence help us to build an effective detection model.

\subsubsection{Phony followers have low social engagement}\label{subsubsection:engagement}
We explore how the purchased follower accounts are connected with their friends. We measure social engagement of users with their friends in form of retweets, @-mentions and favorite count. We also find out language overlap patterns between the users and her friends. 
\paragraph{RTs and @-mentions}
We observed that a large fraction of purchased accounts post only retweets instead of original content. We further explore whether these users retweet the content of their friends or not. If $RT_{count_i}$ is the number of tweets the user has retweeted of her friend $u_i$ and she has $N$ friends, then 
\small 
$$Retweet Ratio = \frac{\frac{RT_{count_i}}{\Sigma_{i=1}^N RT_{count_i}}}{N * RT_{total}}$$
\normalsize
$RT_{total}$ is the total number of retweets done by the user. This \emph{Retweet Ratio} quantifies the number of friends a user has retweeted and the number of times she retweeted them.

\begin{figure}[h!]
  \centering
    \includegraphics[width=0.3\textwidth]{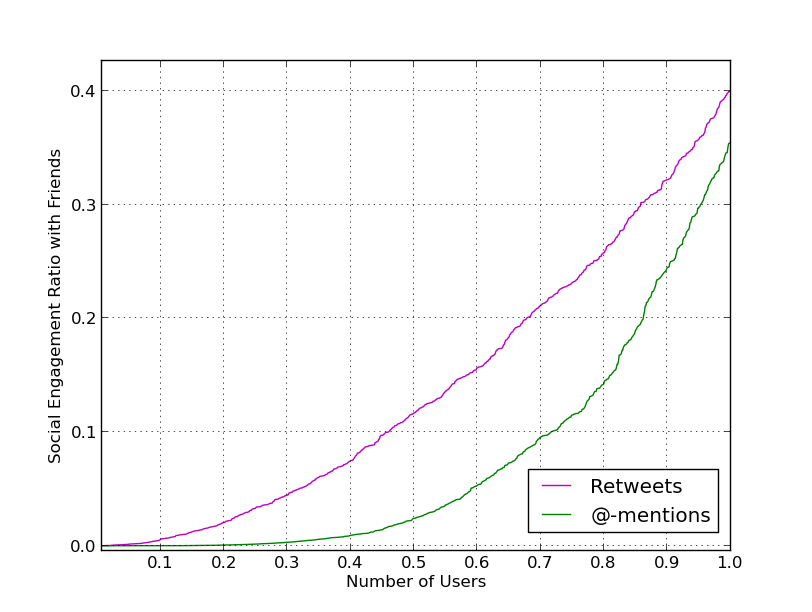}
    \caption{\scriptsize{Social engagement of Purchased Users with their Friends.}}\label{fig:social_engagement}
\end{figure} 

Similarly we define the @-mention ratio to determine whether the user engages in conversations with her friends and to what extent.
\small
$$ @mention Ratio = \frac{\frac{@_{count_i}}{\Sigma_{i=1}^N @_{count_i}}}{N * @_{total}}$$
\normalsize
where $@_{total}$ is the total number of @-mentions by the user. We observe in Figure~\ref{fig:social_engagement} that the highest Retweet Ratio score is 0.45 and the @-mention ratio is 0.35. This shows that though a large fraction of purchased accounts post only retweets, its not the tweets of their friends which they are retweeting. Similarly, low @-mention ratio suggests that purchased followers do not mention their friends. We found the maximum @-mention ratio with the followers of purchased users to be 0.32. This indicates that purchased followers are low quality users and do not engage in conversations with their friends or followers.

\paragraph{Language overlap with Friends and Followers}

\begin{figure}[h!]
  \centering
\subfigure[Distribution]{
        \includegraphics[width=0.19\textwidth]{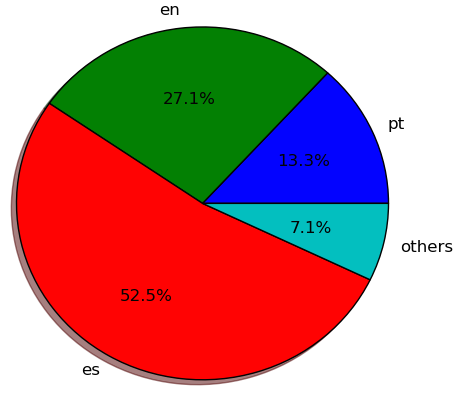}}\label{subfig:language} 
        \subfigure[Number of Languages]{
        \includegraphics[width=0.24\textwidth]{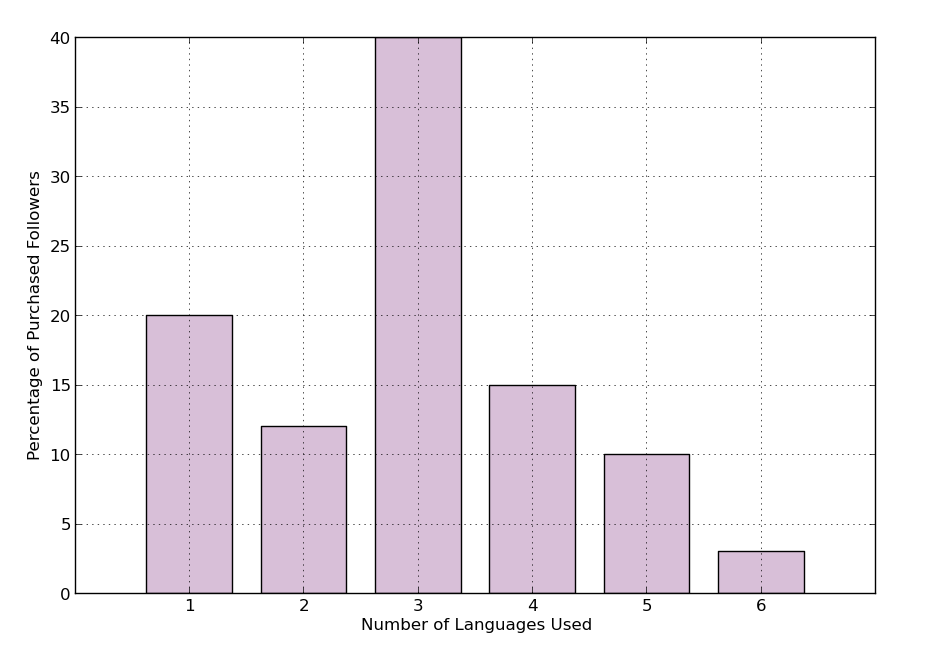}}\label{subfig:NumLang} 
    \caption{\scriptsize{Languages used by Purchased Followers.}}\label{fig:languageAnalysis}
\end{figure}

\begin{figure}[h!]
  \centering
    \includegraphics[width=0.35\textwidth]{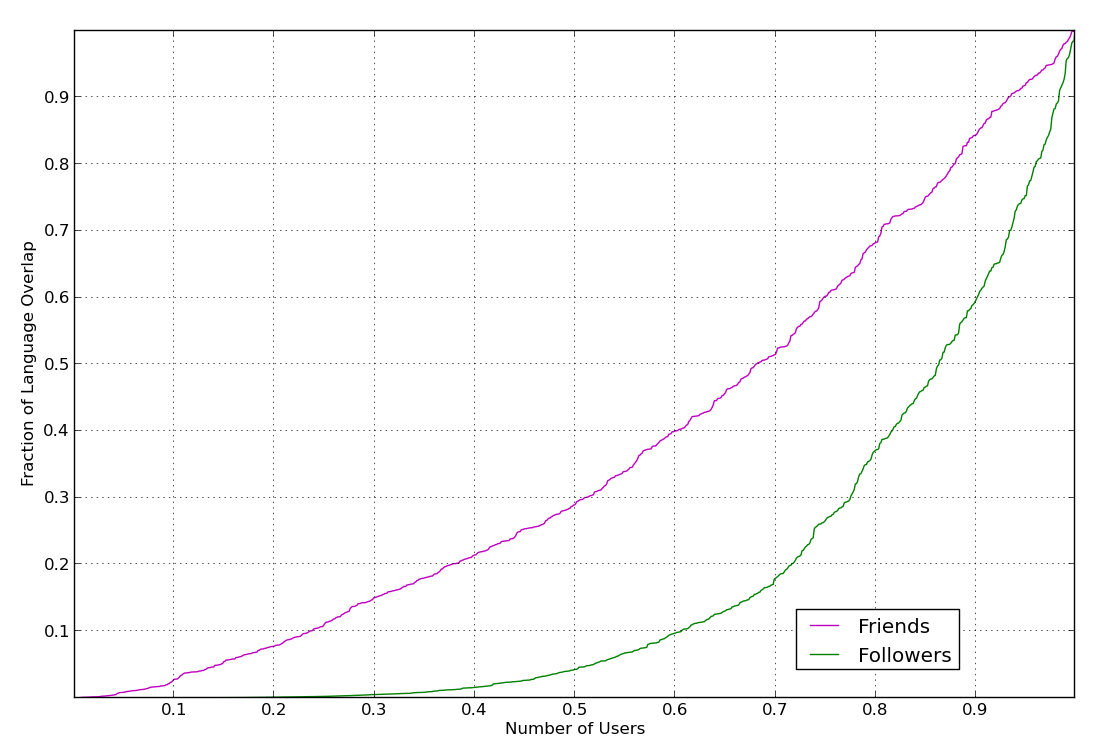}
    \caption{\scriptsize{Language Overlap of Purchased Followers with their Friends and Followers.}}\label{fig:LanguageOverlap}
\end{figure}

We characterize the language used by the purchased followers and the overlap with their friends. Figure~\ref{fig:languageAnalysis} shows the distribution of language of purchased accounts. We observe that a 52\% of the users tweet in spanish. We also found that the purchased followers tweet and retweet in multiple languages as shown in Figure~\ref{fig:languageAnalysis}. Thirteen percent users used 5 or more languages. Only 32\% users posted tweets in less than or equal to two languages. We next find out the overlap of language amongst the purchased accounts with their followers and friends.
Users tweet and retweet in multiple languages. We calculate the \emph{Language Overlap Score} for each user defined as
\small
$$LangOverlap = \frac{\Sigma_{i=0}^N overlap_i}{N}$$
\normalsize
where N is the total number of friends or followers. If $L_f$ is the set of languages used by the friend/followers and $L_u$ is the set of languages used by the purchased user then $overlap_i$ with each friend/follower $u_i$ is defined as
\small
\begin{equation*}
  overlap_i=\begin{cases}
    1, & \text{if $|L_f \cap L_u| \neq 0 $}.\\
    0, & \text{otherwise}.
  \end{cases}
\end{equation*}
\normalsize
We use the \emph{Language Overlap} score to determine how many users tweet in same language as their friends or followers. Figure~\ref{fig:LanguageOverlap} shows that 80\% users had an Overlap Score = 0.37 with their followers and Overlap Score = 0.68 with their friends. This indicates that a large fraction of purchased follower accounts do not care about the content posted by the users they are following. Also, the followers of these users do not have a high language overlap with them.  

\subsubsection{Phony followers have low social reputation}\label{subsubsection:reputation}
\paragraph{Follower-Followee Ratio}
We now look at the relationship between amount of followers and friends for purchased follower accounts. On Twitter, `followers' of a person are the users which subscribe to the posts of that person, i.e., who `follow' her. The `friends' of a person are the users whom she subscribes to. The average number of followers per existing account is 68 and the average number of friends is 60 on Twitter.

\begin{figure}[h]
  \centering
    \includegraphics[width=0.35\textwidth]{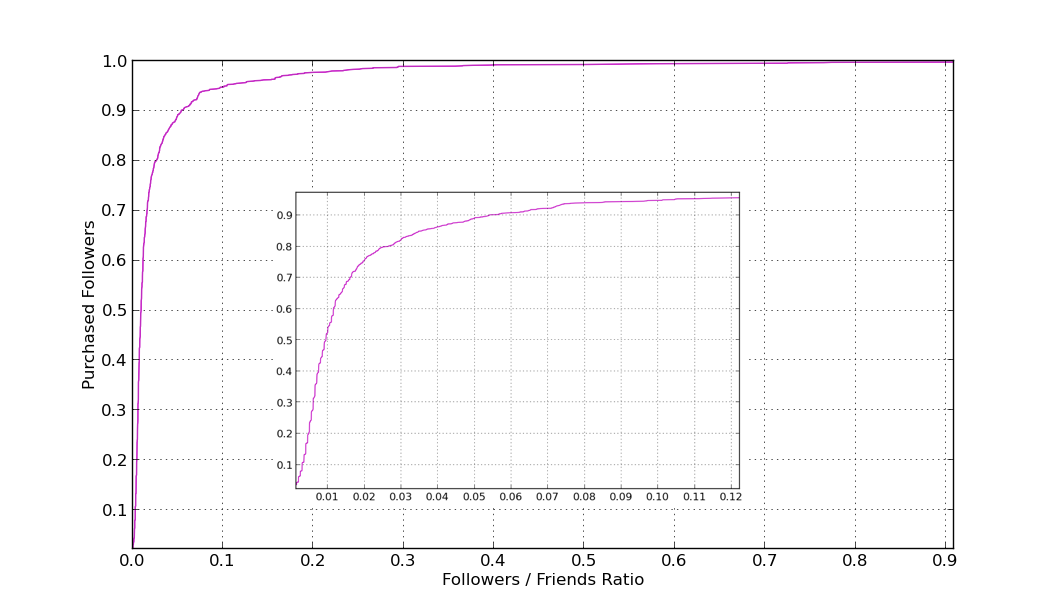}
    \caption{\scriptsize{Follower-Followee ratio of purchased follower accounts.}}\label{fig:follower-followee-ratioCombo}
\end{figure}

Figure~\ref{fig:follower-followee-ratioCombo} shows that the follower/friends ratio fits the power law ($\alpha$ = 1.8209, \emph{error} $\sigma$ = 0.029). We observe that 94\% purchased followers have the follower/friends ratio as only 0.1 and none of the purchased followers had more followers than friends. Low follower/friends ratio indicates that the user does not have a good following, therefore indicating a low social importance.

\begin{figure}[h!]
  \centering
    \includegraphics[width=0.45\textwidth]{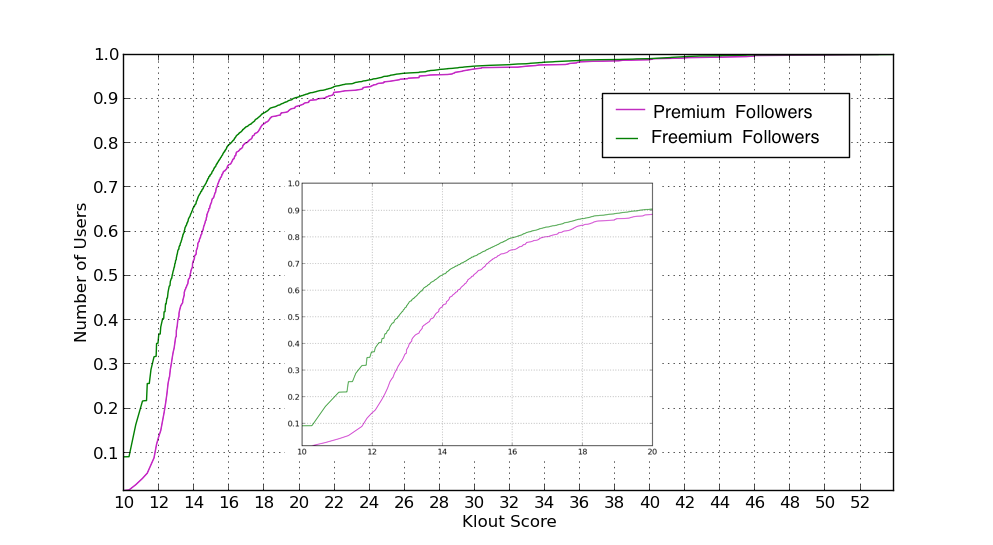}
    \caption{CDF of Klout Score of Purchased Followers.}\label{fig:kloutScoreCombo}
\end{figure}

\paragraph{Klout Score}
To measure the social influence, we use Klout score. We recall that the average Klout score for the social media users is 40. However, as shown in Figure~\ref{fig:kloutScoreCombo}, we found that 90\% of the purchased followers had a Klout score of less than 20. This shows that these accounts do not involve in discussions with other users and have a low influence score.

\subsubsection{Phony followers exhibit high unfollow entropy}\label{subsubsection:entropy}
We found that the purchased follower unfollowed a large number of users regularly. To quantify this behaviour, we calculated the \emph{unfollow entropy} of all the purchased followers. We observed each purchased follower over a span of 15 days and collected her hourly followers. We define normalized \emph{unfollower entropy} $H$ for a user $u_n$ as the following 
\small
$$
H_{u_n} = - \frac{\Sigma_{i=1}^T p_n(f_i) log(p_n(f_i))}{N}
$$
\normalsize
where, $p_n(f_i)$ is the probability that the user $u_n$ will unfollow at time $t_i$. The probability function is defined as
\small
$$
p_n(f_i) = \frac{ucount_i}{\Sigma_{i=1}^T ucount_i}
$$
\normalsize
where $T$ is the number of days for which we monitor the purchased follower and $ucount_i$ is the number of users she unfollowed on $i^{th}$ day. A higher value of unfollow entropy signifies that the user exhibits a suspicious unfollow pattern.

\begin{figure}[h]
  \centering
    \includegraphics[width=0.25\textwidth]{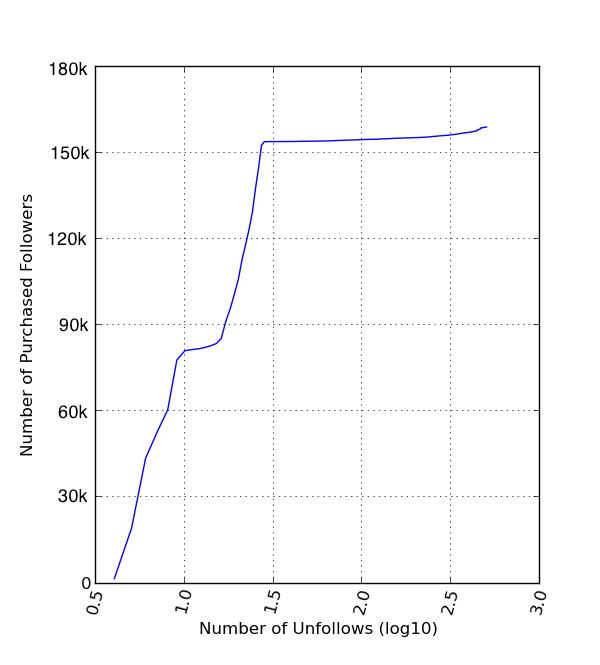}
    \caption{\scriptsize{Unfollow Entropy Rate for Purchased Followers. A large number of followers in our data follow-unfollow their friends multiple times.}}\label{fig:unfollow-entropy}
\end{figure}

Figure~\ref{fig:unfollow-entropy} shows that a large fraction of purchased followers have a high unfollow entropy. The normalized entropy rate for 23\% purchased followers is as high as 0.76 and only 8\% users have a normalized unfollow entropy less than 0.21. To find out whether the users with higher unfollow entropy have lower quality than other users, we compared their normalized unfollow entropy rate with \emph{Klout} score. We found a strong negative correlation (\emph{Pearson correlation coefficient = -0.73}) indicating that users with higher unfollow entropy rate have low social reputation.  

\section{Prediction of Suspicious Following Behaviour}\label{sec:prediction}
In the second part of our study, we build a supervised predictive model to detect suspicious following behaviour on Twitter. In this section, we explain the feature set used for the classification task and the experimental setup.

\subsection{Features for Classification}\label{subsec:features}
For our prediction task to detect suspicious following behaviour, we explore \emph{user profile}, \emph{network}, \emph{content} and \emph{user behaviour} based features. In all, we explore 18 features for our  classification task as described in Table~\ref{table:feats}. \emph{User profile} based features focus upon properties of the Twitter user profile information. The \emph{network} based features describe the relationship of the user with her friends and followers. We next explore the \emph{content} based features to understand the nature of tweets posted by the user and also investigate the \emph{behavioural} features to understand the tweeting patterns and follow dynamics exhibited by the user. For the \emph{network} based features, we constrain our analysis to single hop network of the users due to Twitter API rate limit restrictions. Also, we keep our \emph{content} based analysis limited to  stylistic features of tweets due to the presence of multilingual users in our dataset and the complexity of computation due to transliterated text, misspellings and use of short hand language. Table~\ref{table:feats} enlists all the feature sets we used for our prediction task. 

\begin{table}[h]
\scriptsize
\caption{\scriptsize{Description of the feature sets used for prediction of users with suspicious following behaviour.}}\label{table:feats}
\center
\begin{tabular}{@{}lll@{}}
\toprule
Set & Category & Features \\ \midrule
\multirow{4}{*}{A} & \multirow{4}{*}{User Profile} & presence of bio \\
\multicolumn{1}{c}{} &   & presence of URL in bio \\
\multicolumn{1}{c}{} &   & number of posts \\
\multicolumn{1}{c}{} &   & social reputation \\ \midrule
\multirow{2}{*}{B} & \multirow{2}{*}{Network} & follower / friends ratio \\
 &  & number of followers \\ \midrule
\multirow{6}{*}{C} & \multirow{6}{*}{Content} & hashtags per tweets \\
 &  & spam words used per tweet \\
 &  & length of tweet \\
 &  & number of languages used \\
 &  & number of RTs per tweet \\
 &  & @mentions per tweet \\ \midrule
\multirow{6}{*}{D} & \multirow{6}{*}{Behaviour} & unfollow entropy rate \\
 &  & RT engagement score \\
 &  & @mention engagement score \\
 &  & language overlap \\
 &  & time since last tweet \\
 &  & tweets per day \\ \midrule
\end{tabular}
\end{table}

We explained some of these features in the previous section; here we describe how we calculated the values of remaining features:

\paragraph{\textbf{Presence of bio and URL}} Some Twitter users give description about themselves on their profile which is called \emph{bio}. We check the presence of \emph{bio} for each user under inspection. We also check whether the user has mentioned any external URL in her \emph{bio} and use this as a feature.

\paragraph{\textbf{Social reputation}} We define social reputation by the \emph{Klout} score which gives an estimate of the impact score of the user on various online social networks. 

\paragraph{\textbf{Hashtags per tweet}} We calculate the average number of hashtags used per tweet. We define this metric as
\small
$$
\text{hashtag/tweet} = \frac{\Sigma_{tweet=0}^N \#hashtags}{\#tweets}
$$
\normalsize

\paragraph{\textbf{Spam words used per tweet}} In the earlier section, we noticed that a fraction of purchased follower accounts also spread spam and malicious content. To detect spam in the tweet content, we use a spam word lookup list~\footnote{\url{http://www.mailup.com/spam-words-to-avoid.htm}} and define the following metric
\small
$$
\text{spam words/tweet} = \frac{\Sigma_{tweet=0}^N \text{\#spam words}}{\#tweets}
$$
\normalsize

\paragraph{\textbf{Time since last tweet}} We found that purchased followers exhibiting suspicious following behaviour have very less tweeting activity and are often inactive. To measure  time since the account has been inactive, we find the difference in time in seconds since the latest tweet with the time of our experiment.

These are the discriminative features we use to distinguish between regular and suspicious following behaviour. With the help of these features, we detect users with suspicious follow behaviour in the following section. 

\subsection{Experimental Setup and Classification}\label{subsec:experiment}
For our classification experiment, we consider the 170k public purchased followers as our true positive dataset of suspicious follow behaviour. For the negative class (legitimate follow behaviour), we pick random 170k users from Twitter stream using the streaming API. However, a balanced dataset as ours may create a sample bias. Therefore, to ensure valid results and eliminate the bias, we under-sample our negative class. We draw 10 random but independent subsets from the set of 170k legitimate users (-ve class) and train 10 classifier models based on these 10 subsets along with the 170k samples of the suspicious follow behaviour users (+ve class). We then use 10 fold cross validation and report the average results for our prediction task.

We treat the detection of suspicious follow behaviour as a two class classification problem. In order to detect such behaviour, we use several supervised learning algorithms like Naive Bayes, Gradient Decent, Random Forest etc. However, we achieved highest accuracy and overall best results with \emph{Support Vector Machine} (SVM). 
We use a non-linear SVM with the Radial Basis Function (RBF) kernel for our experiment. Table~\ref{table:svm} gives the details of our experimental setup - dataset description and the parameter values for the SVM classification algorithm. 

\begin{table}[h]
\caption{\scriptsize{Description of the experimental setup for prediction fo suspicious following behaviour.}}\label{table:svm}
\scriptsize
\begin{tabular}{l|l}
\toprule
Dataset & 342,000 users\\
Suspicious (+ve class) & 170,000 users\\
Legitimate (-ve class) & 170,000 users (10 times) \\
Classifier & SVM \\
C & 1,000 \\
alpha & 20.0 \\
Classification Runs & 10 \\
Feature Sets & \{A\}, \{A, B\}, \{A, B, C\}, \{A, B, C, D\} \\
Train-Test Split & 70\%-30\% \\
Cross Validation & 10-fold \\ \bottomrule
\end{tabular}
\end{table}

In order to assess the effectiveness of features, we repeat the classification experiment by incrementally adding each feature set. For evaluation, we used 70-30 split of the training and the testing dataset. We use 10 fold cross validation to report our results. 

\subsection{Classification Results and Evaluation}\label{subsec:classification}
Table~\ref{table:confusion} shows the confusion matrix for our classification task. The confusion matrix defines the percentage of false negatives and false positives. We were able to accurately classify 88.5\% users with suspicious follow behaviour and 89.9\% users with legitimate behaviour. This shows that we are able to detect suspicious following behaviour to a good extent. For the evaluation of our classification result, we used the standard evaluation metrics -- accuracy, F-measure and Area under the Curve (AUC).

\begin{table}[h]
\center
\caption{\scriptsize{Confusion Matrix -- Classification Results of distinguishing legitimate users from those exhibiting suspicious following behaviour.}}\label{table:confusion}
\scriptsize
\begin{tabular}{@{}cccc@{}}
\toprule
 &  & \multicolumn{2}{c}{Predicted} \\ \midrule
 &  & Suspicious & Legitimate \\
\multirow{2}{*}{True} & Suspicious & 88.5 & 11.5 \\
 & Legitimate & 9.7 & 89.9 \\  \bottomrule
\end{tabular}
\end{table}

\begin{figure}[h]
  \centering
    \includegraphics[width=0.4\textwidth]{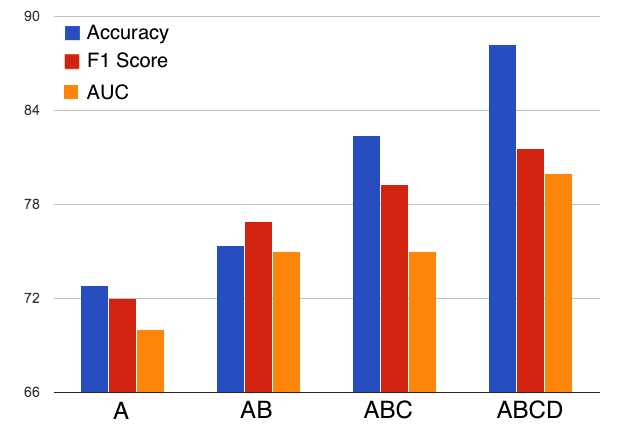}
\caption{\scriptsize{Classification accuracy to predict suspicious following behaviour on incremental feature addition.}}\label{fig:incremental}
\end{figure}

As discussed in the previous section, we incrementally added feature sets to evaluate the effectiveness of all the features. Figure~\ref{fig:incremental} shows the performance of our classifier on Accuracy, F1 score and AUC metrics when feature sets are incrementally added. We see that each feature set has a positive effect on the performance of the classifier across all metrics. We also observed that adding behavioural based features suddenly increase the overall accuracy of our classification model. We received a maximum accuracy of 89.2\%.

\subsection{Feature Importance}\label{subsec:importance}
In this section, we look at the importance of features used for suspicious follow behaviour detection. We found that behavioural features are important to detect suspicious behaviour. \emph{Unfollow entropy} rate plays an important role; it is defined as the frequency with which the user is unfollowing her friends over time. Some of the most informative features we received after our classification task were \emph{unfollow entropy, RT-engagement ratio, @mention-engagement ratio, Language Overlap and Social Reputation}. The other informative and discriminative features were the use of multiple hashtags and spam words in the tweets. The user profile based features were the least helpful in detection of suspicious follow behaviour. One possible reason for this could be that a large fraction of legitimate users do not add a \emph{bio} or engage in heavy conversations on Twitter.

\section{Discussion}\label{sec:discussion}
Follower markets enable users to artificially boost their percieved reputation and popularity. Social media usually do not have any defined metrics to gauge the influence of its users. Therefore, the number of connections and following of a user is believed to be a true reflection of user's influence and popularity. Our technique can help the social media operator to identify accounts exhibiting suspicious following behavior by reducing their search space for synthetically created accounts by merchants. Suspension or deletion of such synthetic accounts can effectively disrupt the services of underground market.

Our study can also be utilized by legitimate users who want to maintain a social network profile with only trustworthy connections. Phony followers controlled by merchants often follow few random users to maintain their follower-friends count and hence evade detection. However, legitimate users who want to avoid and block such users from following them can use our technique to identify their own followers with suspicious behavior with a high accuracy. Such measures can be used against scams like [3] where hoaxters bought 75k fake followers for a legitimate user. In case of such attacks, our mechanism can help the victim user by identifying fake followers and removing them from user's connections.

Our study also sheds light on the source of this problem, i.e., the follower \emph{merchants}. Follower merchants are responsible for the generation of 10-20\% spam accounts that exist on Twitter~\cite{thomas2013trafficking}. Therefore, its important to understand the dynamics of these merchants. We show that though there exist multiple merchants, the follower market has an \emph{oligopolist} structure; there exist only few market leaders attracting most of the customers. Mitigating the operations of market leaders can potentially reduce the infiltration of synthetic and compromised accounts into the social network.

The merchants may try to adapt their operations once we are effectively able to detect accounts exhibiting suspicious following behaviour. However, one of our key observations is that we can identify the market leaders. Currently, there exists a good amount of difference between the market share of leaders and other merchants. Hence, even if the merchants evolve their techniques, the market leaders would not be difficlt to distinguish. Bringing down these market leaders can help the social media operators to a large extent. Secondly, we observe that accounts with suspicious following behaviour have a high unfollow entropy rate and keep unfollowing and following back their friends. Since the merchants try to provide bulk followers in a short amount of time, they would have to manage who-to-follow of their synthetic and compromised accounts to effectively cater to their customers. These phony followers provided by merchants would either get detected because of our technique, or the merchants would have to slow down their operations by reducing the follow-unfollow activity of the accounts they control. Slowing down merchant's activities would indirectly benefit the social media operator because the merchants will not be able to cater to a larger customer base. However, this argument holds if merchants can create only limited number of synthetic profiles. In case merchants have a large number of sythetic profiles with a good follower-friends ratio with low unfollow entropy, then they will be able to evade detection by out technique. However, creation of a large number of sythetic accounts will have a higher cost to the merchants and can also be detected by already existing techniques. 

\section{Limitations and Conclusion}\label{sec:conclusion}
In this study, we present a landscape of Twitter underground follower market. We focus of the building blocks of follower market - \emph{merchants}, \emph{customers}, \emph{phony followers}. We use a dataset of 60 freemium and 57 premium merchants for our study which are most responsive and active. We collect 170k phony followers from these merchants. Also, we took hourly snapshots of the merchant websites which can be further improved to collect a larger dataset. Though there exist millions of compromised and synthetic accounts controlled by the follower merchants, our dataset covers a large portion of underground market over a course of 4 months. Therefore, we posit that our results would be scalable over a much larger network of underground merchants. 

To summarise, we present the following in this study (i) We measure the quality of service and uncover the underlying hierarchy of merchants. We discover an \emph{oligopoly} structure of the merchants, (ii) We analyze the reputation and profile attributes of the market customers to understand who they are are which merchants they subscribe to, (iii) We study suspicious following behaviour of the phony followers and build a supervised learning model to distinguish them from legitimate users. This is the first study to landscape all aspects of an underground market to understand the underlying structure and characteristics. 

\bibliographystyle{abbrv}
\bibliography{Freemium}  
\balancecolumns
\end{document}